\documentclass[prb,showpacs,superscriptaddress,titlepage,amsmath,amssymb,twocolumn]{revtex4-1}

\usepackage{graphicx}
\usepackage{bbold}
\usepackage{epsfig}
\usepackage{subfigure}
\usepackage{hyperref}
\usepackage{epstopdf}
\usepackage{dcolumn}
\usepackage{bm}
\usepackage{color}
\usepackage{amsmath}      
\usepackage{amssymb}   

\begin{document}

\title{
An SU(3) Yang-Mills Structure for Electron-Phonon Interactions Resulting from Strong Electron Correlations in 2D hexagonal lattices
}
\author{J.~M.~Booth}
\email{jamie.booth@rmit.edu.au}
\affiliation{ARC Centre of Excellence in Exciton Science, RMIT University, Melbourne, Australia}

\date{\today}

\begin{abstract}
Concise and powerful mathematical descriptions of the interplay of spin and charge degrees of degrees of freedom with crystal lattice fluctuations are of extreme importance in materials science. Such descriptions allow structured approaches to optimizing material efficiencies resulting in considerable resource savings and higher performance devices. In this work, by re-imagining the the Gell-Mann matrices as 3$\times$3 linear transformations acting on a column vector of position states, an SU(3) theory of the interplay between lattice fluctuations and strong electron correlations in 2-dimensional hexagonal materials such as graphene is formulated. 
\end{abstract}

\maketitle
\section{Introduction}
There is no doubt that two-dimensional hexagonal materials such as graphene are of enormous significance in contemporary materials science. Graphene in particular exhibits many useful electronic phenomena, and in addition has attractive theoretical qualities. It exhibits extremely high electron mobility at room temperature, and can sustain high densities of electric currents\cite{Novoselov2012}. It also has a high Young's modulus and can be readily chemically functionalized\cite{Novoselov2012}.

Magic Angle twisted bilayer Graphene superlattices have been found to exhibit both correlated insulating behavior\cite{Cao2018_2} and unconventional superconductivity\cite{Cao2018} with a T$_{c}$ of up to $\sim$ 1.7 K. A model for the metal-insulator transition in graphene superlattices was proposed which took the form of a two-orbital Hubbard Model on an emergent honeycomb lattice\cite{Yuan2018}.

In a previous study\cite{Booth2020} it was found that an SU(2) Yang-Mills description of electron-phonon interactions in linear systems such as vanadium dioxide can be developed by assuming that the transverse phonons couple to the electron spin via a Rashba-type mechanism, while charge ordering which leaves the spins unaffected is carried out by the longitudinal modes which constitute a ``neutral" boson. The SU(2) interaction vertex described there has many advantages over the standard U(1) approach to electron-phonon coupling. It contains both charge and spin-ordering and manifests at neighboring atomic sites and therefore can describe phase transitions in which spin-ordering is also present.

In a subsequent study\cite{Booth_Wilson_2020} it was shown that the structure of the SU(2) vertex arises from minimizing electron correlations. That is, the presence of strong electron correlations, and thus electron physics obeying the Hubbard Model, can generate phonons which are matrix operators acting on more than one nucleus simultaneously. In that study a simple Hubbard Hamiltonian was used:

\begin{equation}
	H=-t\sum\limits_{\langle ij\rangle}^{}(c^{\dagger}_{i\sigma}c_{j\sigma}+c^{\dagger}_{j\sigma}c_{i\sigma})+\\U\sum\limits_{i}n_{i\uparrow}n_{i\downarrow}
	\label{HubbardH}
\end{equation}

where $t$ is the electron hopping energy, and $U$ is the usual on-site repulsion term which penalizes double occupancies. The phonons identified in that study order the charge and spin in such a manner as to reduce the energy of the electrons.

In quasi-linear systems such as vanadium dioxide it was shown that the matrix valued electron phonon vertex was a 2$\times$2 linear transformation parametrized by the Pauli matrices: an SU(2) gauge theory. 

One of the biggest advantages of this approach was that by identifying the phonons which correspond to charge and spin fluctuations which lower the electron energy, the system near T$_{c}$ can be described by an interacting liquid of SU(2) bosons. This description can in turn be simply modelled using an Ising-type Hamiltonian which leads to the formation of a phase coherent phonon state which breaks the symmetry and opens a gap\cite{Booth_Wilson_2020}.

In this work we repeat the same process, but we examine the SU(3) gauge group and find that it also describes charge and spin fluctuations in a low dimensional structure, however unlike linear systems as per SU(2) it describes 2D hexagonal systems, such as graphene. Importantly, we assume the same Rashba-type mechanism is active which couples the spin raising and lowering operators to polarization vectors with components in the $x$ and $y$ directions, while polarizations along the $z$ axis are neutral.

While the Hubbard Model used here describes insulator to metal transitions, the main point of this work is to show that the electron charge and spin fluctuations of a hexagonal system described by the Hubbard Model can be mapped to an SU(3) gauge theory. Therefore the correlated metallic state is an interacting liquid of SU(3) bosons. 

The properties of this liquid are then given by the scattering amplitudes of the SU(3) bosons. Thus the strongly correlated electron system is mapped to a Pure Yang-Mills theory: the interactions of the bosons contain a description of both the electron and the lattice behavior.

This discussion presents the algebraic formalism of the SU(3) approach by first presenting a review of the SU(2) formalism\cite{Booth2020,Booth_Wilson_2020}. This is then extended to SU(3) by identifying the eight generators of the group with collective phonon modes which act on three site sub-units of the hexagons. It is shown that like the SU(2) group of linear systems (e.g. VO$_{2}$), the SU(3) generators correspond to atomic motions which minimize the energy of a system described by the Hubbard Model of Equation \ref{HubbardH}.

The calculation of boson-boson scattering terms is not included in this discussion. It requires the formal machinery of Yang-Mills scattering amplitudes to be adapted to a non-relativistic environment. Gluon scattering amplitudes are highly non-trivial due to the number of generators, and the redundancy built into the diagrammatic gauge theory formalism.\cite{Dixon2013a,Arkani-Hamed2017} This aspect of the approach is thus left to a future study.

\section{SU(2) Gauge Theory Review}
The key to viewing crystalline materials through the lens of non-Abelian gauge theory is the coupling of the spinors to the atomic motions through the Pauli matrices, and re-imagining the gauge transformations as linear transformations acting on position states. 

The tree-level form of a non-Abelian interaction vertex is given most generally by\cite{Zee2003}:
\begin{equation}
\bar{\psi}\gamma^{\mu}\hat{G}_{\mu}^{a}\hat{T}^{a}\psi
\label{vertex}
\end{equation}
While unfamiliar for condensed matter physicists, its application to condensed matter systems can be broken up into simple parts. The electron states $\psi$ are a column vector of position states, the number of which is given by the dimension of the SU(N) group, i.e. for an SU(2) group there are two position states. In the case of vanadium dioxide these are neighboring vanadium atoms on the one-dimensional vanadium chains of the structure. Explicitly:
\begin{equation}
\psi = \begin{pmatrix}
\psi(x_{i})\\\psi(x_{j})
\end{pmatrix}
\end{equation}
where $x_{i}$ and $x_{j}$ are neighbouring atomic sites on the chain. 

Each position state electron wavefunction carries spinor indices, and the gamma matrices $\gamma^{\mu}$ act on these spinor variables. Importantly each $\psi(x_{i})$ is a \textit{four} component spinor (examples of which are the Nambu spinors\cite{Booth2020}), as it includes both electron and hole spinor variables, which necessitates the use of gamma matrices.  These are double-stacked Pauli matrices, with a sign inversion ensuring the hole states satisfy the Weyl equation. They are given by:
\begin{equation}
\gamma^{0} = \begin{pmatrix}0&\mathbb{1}\\\mathbb{1}&0\end{pmatrix}, \gamma^{i} = \begin{pmatrix}0&\sigma^{i}\\-\sigma^{i}&0\end{pmatrix}
\end{equation}
where $\mathbb{1}$ is the 2$\times$2 unit matrix, and the $\sigma^{i}$ are the usual Pauli matrices. 

The $\hat{G}^{a}_{\mu}$ are a set of $N^{2}-1$ vector fields, which are quantized in the usual manner (note these may have non-linear dispersion, unlike Lorentz invariant vector fields):
\begin{equation}
\hat{G}^{a}_{\mu}(x) = \int \frac{d^{3}\mathbf{p}}{{2\pi}^\frac{3}{2}2E_{\mathbf{p}}^{\frac{1}{2}}}\sum_{\lambda}\big[\hat{a}_{\mathbf{p}}\epsilon^{\lambda}_{\mu}(p)e^{ipx} + \hat{a}_{\mathbf{p}}^{\dagger}\epsilon^{*\lambda}_{\mu}(p)e^{-ipx}\big]
\label{A-field}
\end{equation}
where $a$ (no hat) labels the vector field, and these multiply the $\hat{T}^{a}$, which are the generators of the gauge group; $N\times N$ matrices. In the SU(2) theory developed for quasi-linear systems, the generators of the SU(2) group are the Pauli matrices. However, as stated above the key to the utility of the Yang-Mills formalism for crystal systems is in re-imagining these generators as linear transformations which act on position states.

What this means in practice is that the generators take the polarization vector from the field $\hat{G}^{a}_{\mu}$ and apply it to $N$ sites simultaneously, with coefficients given by the generator $\hat{T}^{a}$. The simplest illustrative example of this would be $\hat{T}^{3}$ of the SU(2) theory, contracted with a vector field (omitting the gamma matrices for clarity):
\begin{equation}
\hat{W}^{3}_{\mu}\sigma^{3}\psi = \begin{pmatrix}\hat{W}^{3}_{\mu}&0\\0&-\hat{W}^{3}_{\mu}\end{pmatrix}\begin{pmatrix}\psi(x_{i})\\\psi(x_{j})\end{pmatrix}
\end{equation}
This operation applies the polarization vector $\epsilon^{\lambda}_{\mu}$ (with the appropriate creation of annihilation operator) of the field $\hat{W}^{3}_{\mu}$, to the positions of the nuclei to which the electron wavefunctions are bound in the column vector. However with a crucial change of sign for $\psi(x_{j})$, as the generator $\hat{T}^{3}$ is just the Pauli matrix $\sigma^{3}$. Therefore, this interaction describes equal and opposite motions of neighboring atoms, a Peierls pairing.

This describes the charge ordering, however the gamma matrices also include spin ordering. For this to manifest in a convenient form we must choose a coordinate system in which the diagonal elements of the generators act on the $z$-components of the position states, while the off-diagonal terms act on the $x$ and $y$ components. In particular, if we re-label the Pauli matrices $\sigma^{1}, \sigma^{2}, \sigma^{3}\rightarrow \sigma^{x}, \sigma^{y}, \sigma^{z}$, then the SU(N) generators which are derived from the Pauli matrices must multiply the corresponding component of the boson polarization vector. When this is contracted with the corresponding gamma matrix, the correct spin and charge ordering occurs. 

Again, taking SU(2) as an example, the first generator is $\sigma^{1} = \sigma^{x}$. Choosing a coordinate system such that the polarization vector $\epsilon^{1}_{\mu}$ of the corresponding boson field $\hat{W}^{1}_{\mu}$ has only one non-zero component, the $x$ component, and writing $\gamma^{1} = \gamma^{x}$ gives:
\begin{equation}
\bar{\psi}\gamma^{\mu}\hat{W}_{\mu}^{a}\hat{T}^{a}\psi = \bar{\psi}\gamma^{x}\hat{W}^{1}_{x}\sigma^{x}\psi
\end{equation}
The operator sandwiched between the electron states is:
\begin{equation}
\gamma^{x}\hat{W}^{1}_{x}\sigma^{x} = \begin{pmatrix}0&\epsilon_{x}\begin{pmatrix}0&\sigma^{x}\\-\sigma^{x}&0\end{pmatrix}\\\epsilon_{x}\begin{pmatrix}0&\sigma^{x}\\-\sigma^{x}&0\end{pmatrix}&0\end{pmatrix}
\end{equation}

\quad\\

\begin{figure}
	\includegraphics[width=\columnwidth]{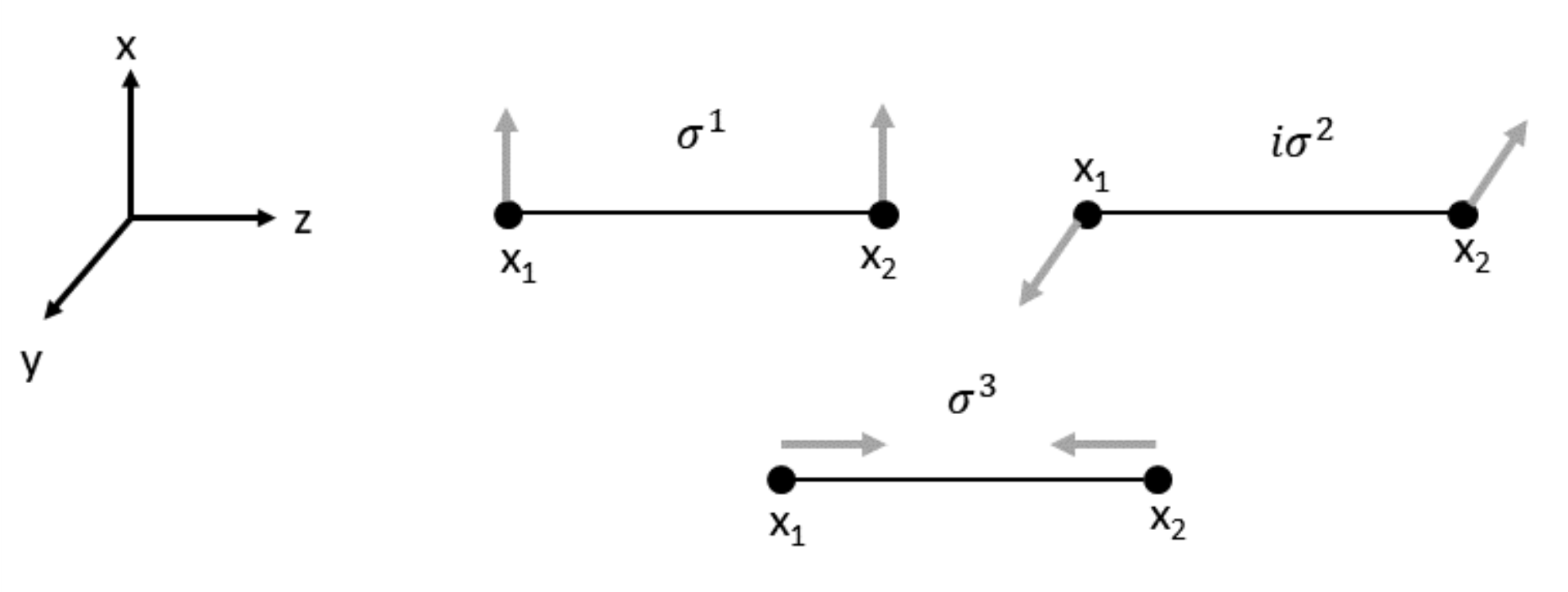}
	\caption{\raggedright{Atomic motions which correspond to the different SU(2) modes, $\hat{W}^{1}$, $\hat{W}^{2}$ and $\hat{W}^{3}$.}}
	\label{SU2}
\end{figure}

The term in the first column applies a displacement to the nucleus position of $\psi(x_{i})$, and operates on the electron spinor variables with $\sigma^{x}$ and on the hole spinor variables with $-\sigma^{x}$. The term in column two applies the same operation to the nucleus position of $\psi(x_{j})$. The generator has only positive components so the displacement vectors are in the same direction, and this is depicted schematically in the first diagram of Figure \ref{SU2}. 

The interaction vertex for the generator corresponding to $\sigma^{y}$, with $\gamma^{2}$ relabeled as $\gamma^{y}$, is obtained in the same manner:
\begin{equation}
\gamma^{y}\hat{W}^{2}_{y}\sigma^{y} = \begin{pmatrix}0&\epsilon_{y}\begin{pmatrix}0&-i\sigma^{y}\\i\sigma^{y}&0\end{pmatrix}\\\epsilon_{y}\begin{pmatrix}0&i\sigma^{y}\\-i\sigma^{y}&0\end{pmatrix}&0\end{pmatrix}
\end{equation}
\quad
where the imaginary units which of $\sigma^{y}$ have been grouped with the Pauli matrices which make up $\gamma^{y}$. Comparing this with the result for the $\sigma^{x}$ generator, we see that we can make linear combinations of these two vertices which correspond to spin raising and lowering operators:

\begin{widetext}
\begin{multline*}
\frac{1}{\sqrt{2}}\gamma^{\mu}\big(\hat{W}^{1}_{\mu}\sigma^{1}+\hat{W}^{2}_{\mu}\sigma^{2}\big) = \\
\begin{pmatrix}0&\frac{1}{\sqrt{2}}\bigg(\epsilon_{x}\begin{pmatrix}0&\sigma^{x}\\-\sigma^{x}&0\end{pmatrix}+\epsilon_{y}\begin{pmatrix}0&-i\sigma^{y}\\i\sigma^{y}&0\end{pmatrix}\bigg)\\\frac{1}{\sqrt{2}}\bigg(\epsilon_{x}\begin{pmatrix}0&\sigma^{x}\\-\sigma^{x}&0\end{pmatrix}+\epsilon_{y}\begin{pmatrix}0&i\sigma^{y}\\-i\sigma^{y}&0\end{pmatrix}\bigg)&0\end{pmatrix}\\ = \begin{pmatrix}0&\begin{pmatrix}0&\hat{S}^{-}\\-\hat{S}^{-}&0\end{pmatrix}\\\begin{pmatrix}0&\hat{S}^{+}\\-\hat{S}^{+}&0\end{pmatrix}&0\end{pmatrix}
\end{multline*}
\end{widetext}
where we have set $\epsilon_{x} = \epsilon_{y} = 1$ in the last line. When acting on the column vector of spinor states this will flip the spins of neighboring electrons into an antiferromagnetic order. This vertex, in which equal polarization vectors in the $x$- and $y$-directions when contracted with gamma matrices gives antiferromagnetic ordering mirrors the ordering seen in the M$_{2}$ structure of vanadium dioxide.\cite{Pouget1974} The M$_{2}$ form contains two alternating chain structures of vanadium atoms. One of which pairs along the collinear axis, while the other orders antiferroelectrically. The antiferroelectric ordering also coincides with anftiferromagnetic ordering of the spins. This vertex describes a combination of antiferromagnetic and antiferroelectric ordering, as required.

The third SU(2) generator describes Peierls pairing, however it leaves the spins unchanged as it is contracted with the $\gamma^{3}$ matrix, which is comprised on $\sigma^{z}$ matrices. Re-labeling $\gamma^{3} = \gamma^{z}$ and $\sigma^{3} = \sigma^{z}$:

\begin{equation}
\gamma^{z}\hat{W}^{3}_{z}\sigma^{z} = \begin{pmatrix}\epsilon_{z}\begin{pmatrix}0&\sigma^{z}\\-\sigma^{z}&0\end{pmatrix}&0\\0&-\epsilon_{z}\begin{pmatrix}0&\sigma^{z}\\-\sigma^{z}&0\end{pmatrix}&0\end{pmatrix}
\end{equation}

Thus these three vertices which are derived from the three generators of the SU(2) gauge group describe antiferroelectricity, Peierls pairing and antiferromagnetic ordering when acting on a two component column vector of neighboring spinor states. 

In a previous work it was shown that the polarization vectors which correspond to the SU(2) bosons are rather simply related to electron correlations arising from the Hubbard Model\cite{Booth_Wilson_2020}. This combination of antiferromagnetic ordering and Peierls pairing will lower the energy of the electron liquid, and therefore the Hubbard Model physics is the source of these SU(2) bosons.

As vanadium dioxide undergoes a metal-insulator transition which coincides with a crystal structure transformation, the interest in that system is on vacuum expectation values of the SU(2) bosons as the system passes through T$_{c}$. However, above the critical Temperature, the  tetragonal VO$_{2}$ form will be a liquid of strongly interacting electrons which are generating SU(2) lattice fluctuations. 

\section{2d Hexagonal Lattices and SU(3)}
The question then becomes, if pairing and antiferromagnetic spin ordering lower the electron correlations, can this formalism which intrinsically contains such phenomena, be used to describe systems which are not simple linear chains? The answer, at least for strongly correlated 2D hexagonal systems, is yes. However, there is a slight generalization needed, and the gauge group must be expanded to SU(3).

The Gell-Mann matrices which parametrize the SU(3) gauge group of Quantum Chromodynamics are given by:
\begin{multline}
\lambda_{1} = \begin{pmatrix}0&1 &0\\1&0&0\\0&0&0\end{pmatrix},\quad \lambda_{2} = \begin{pmatrix}0&-i&0\\i&0&0\\0&0&0\end{pmatrix}, \quad \lambda_{3} = \begin{pmatrix}1&0&0\\0&-1&0\\0&0&0\end{pmatrix}\\ \lambda_{4} = \begin{pmatrix}0&0&1\\0&0&0\\1&0&0\end{pmatrix}, \quad \lambda_{5} = \begin{pmatrix}0&0&-i\\0&0&0\\i&0&0\end{pmatrix}, \quad \lambda_{6} = \begin{pmatrix}0&0&0\\0&0&1\\0&1&0\end{pmatrix}\\ \lambda_{7} = \begin{pmatrix}0&0&0\\0&0&-i\\0&i&0\end{pmatrix}, \quad \lambda_{8} = \frac{1}{\sqrt{3}}\begin{pmatrix}1&0&0\\0&1&0\\0&0&-2\end{pmatrix}
\end{multline}
The key to understanding the physics of applying this approach to condensed matter systems is identifying the three spin operators (i.e. the Pauli matrices) which come from the gamma matrices, with the coefficients which arise in the gauge group generators. 

In the Gell-Mann matrices above, the interaction vertex multiplies each of the entries in the matrix with a gamma matrix. The gamma matrices themselves are comprised of the Pauli matrices which give the required spin ordering.

However, the Pauli matrices are tied to the polarization vectors of the bosons as in the SU(2) theory above. So, to generate pairing which is a ``neutral" mode, i.e. doesn't affect the spin, we need the Gell-Mann matrix to contain $1$ and $-1$ for the paired polarization vectors, and for this to be contracted with $\gamma^{3}$, which we re-iterate is:
\begin{equation}
\gamma^{3} = \begin{pmatrix}0&\sigma^{z}\\-\sigma^{z}& 0\end{pmatrix}
\end{equation} 
Therefore the polarization vectors of this mode must be aligned along the $z$-axis, in order for the spin operators to correspond to the coordinate axes of the crystal.

To apply the spin raising and lowering operators to the structure we need matrices which contain only terms equal to $1$ to contract with $\sigma^{x}$ which comes from $\gamma^{1}$, and separate generators which contain $\pm i$ to contract with $\gamma^{2}$ which contains $\sigma^{y}$ terms, to use in linear combinations which give the spin raising and lowering operators. However these must correspond to polarization vectors in the $x$, and $y$ directions.

Therefore, the coordinate system is heavily constrained. Thankfully a coordinate system can be mapped to the SU(3) generators, and it describes a 2D hexagonal structure. 
\begin{figure}
\includegraphics[width=0.8\columnwidth]{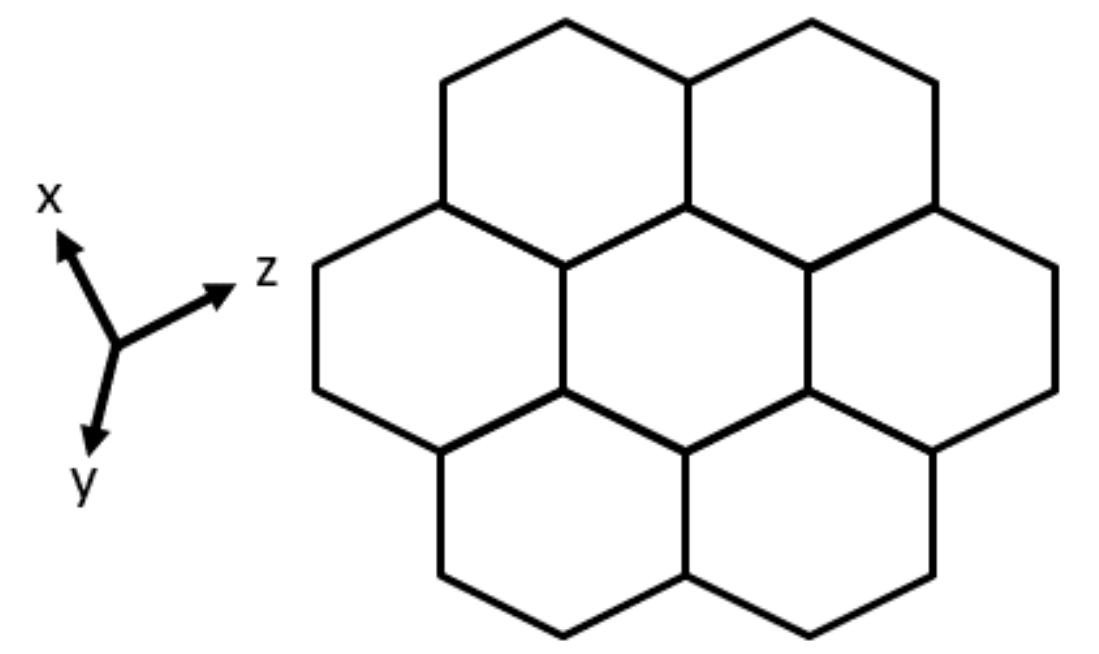}
\caption{Schematic of the lattice structure of a two-dimensional hexagonal system such as graphene, and the coordinate system used in this work. The system is Cartesian; the $z$-axis aligns with two of the six sides of each hexagon, and the $x$- and $y$-axis rotation with respect to this is a free parameter determined by experiment.}
\label{Graphene}
\end{figure}
Figure \ref{Graphene} gives a schematic illustration of a two-dimensional hexagonal structure with the corresponding coordinate axes. The $z$-axis aligns with two sides of each hexagon, while the $x$- and $y$- axes are orthogonal, with the rotation with respect to the $z$-axis determined by experiment.

By orienting the coordinate axes in this manner, a remarkable coincidence between the pairing motions and the diagonal generators of the SU(3) group appears. As stated above the pairing modes must contain entries in their respective generators which are opposite in sign. Obviously, only $\lambda_{2}, \lambda_{3}, \lambda_{5}, \lambda_{7}$, and $\lambda_{8}$ qualify. However, $\lambda_{2},\lambda_{5}$ and $\lambda_{7}$ contain imaginary units $\pm i$ needed for spin raising and lowering. 

There is also the issue that as the structure is not linear, it is difficult to write down generators that pair more than two of the six atoms in each ring using polarization vectors which have only a non-zero $z$-component.

\begin{figure}
	\includegraphics[width=\columnwidth]{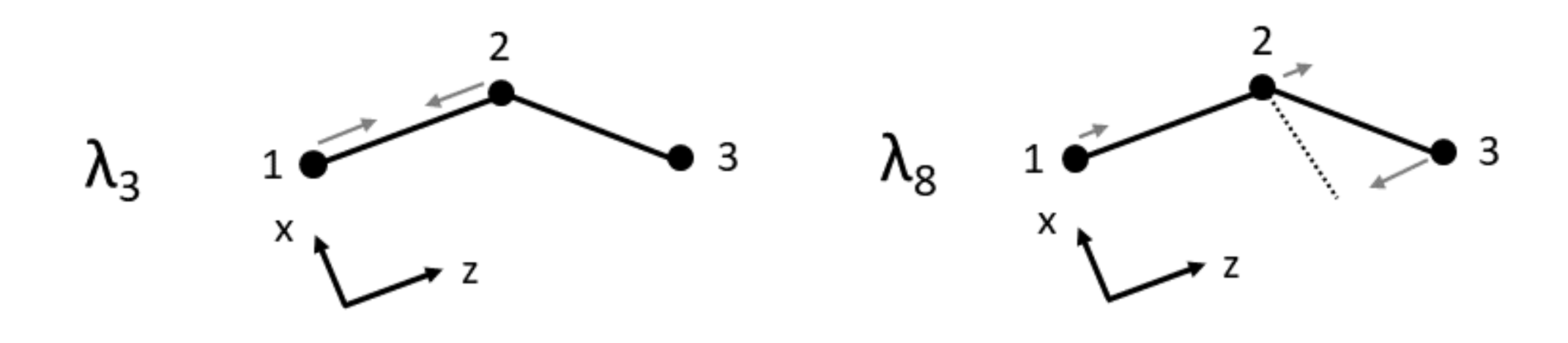}
	\caption{Schematic of the actions of the $\lambda_{3}$ and $\lambda_{8}$ generators when applied to the three-atom unit.}
	\label{z-modes}
\end{figure}

Figure \ref{z-modes} shows how this can be done. The first step is to break each 6 atom ring into two chains of three atoms, an upper chain and a lower chain. This gives the required three component column vector of nuclei position states for the generator to act on. Then, looking at the action of $\lambda_{3}$ and orienting the $z$-axis as shown, we see that like the $\sigma^{3}$ generator of the SU(2) theory, it pairs the first two neighboring atoms on the chain.

\begin{figure}
	\includegraphics[width=\columnwidth]{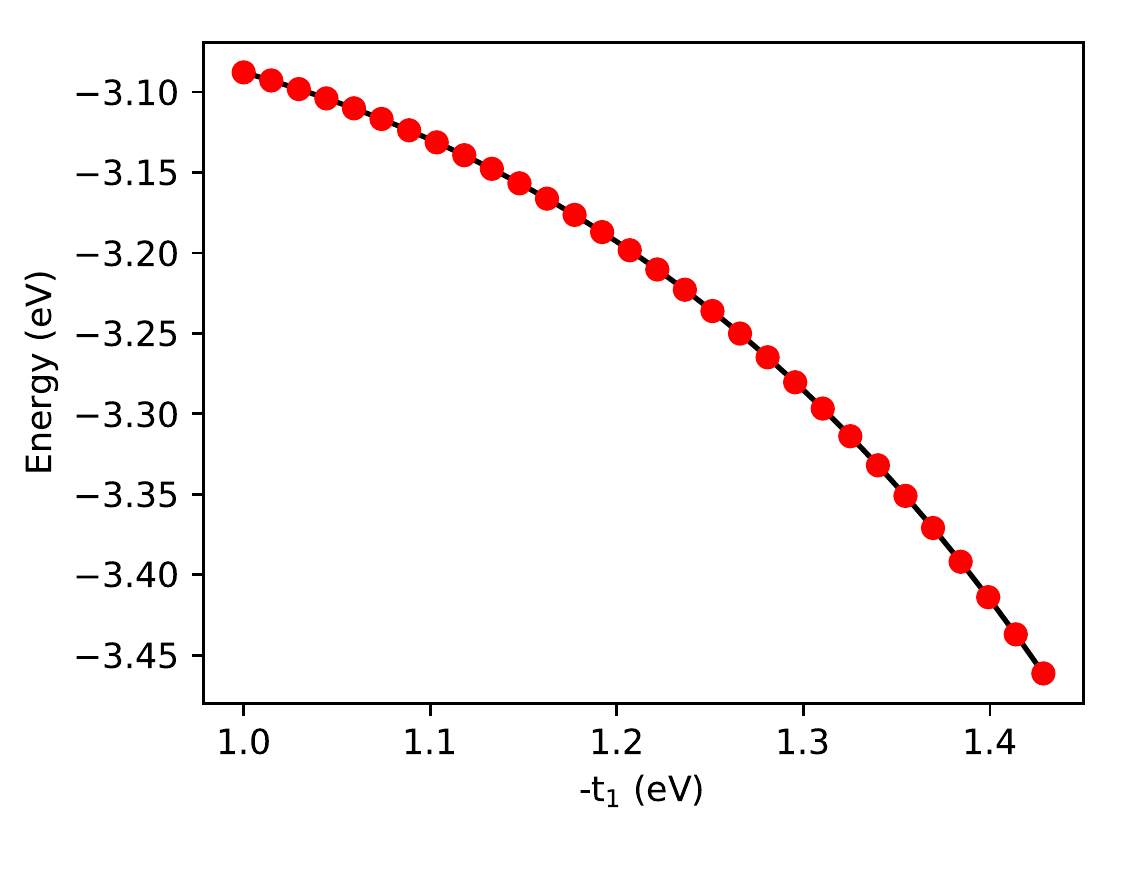}
	\caption{Plot of the energy of a hexagonal Hubbard system in which the first two atoms are paired, corresponding to the generator $\lambda_{3}$, with the hopping energy between atoms 1 and 2 (denoted by -t$_{1}$) decreasing from -1.0 eV to -1.44 eV. The on-site energy $U$ is constant at 5 eV.}
	\label{t1}
\end{figure}

Figure \ref{t1} illustrates the energy of a single hexagonal system in which the electron part of the Hamiltonian is given by Equation \ref{HubbardH}, and the first two atoms of the hexagon are paired, thereby decreasing the hopping energy $t_{1}$ between them, corresponding to the $\lambda_{3}$ diagram of Figure \ref{z-modes}, with the on-site energy term constant at $U$ = 5 eV.  

Despite this mode only pairing two of the six atoms of the hexagon, a substantial decrease in the electronic energy of the system is observed despite the corresponding increases in hopping energies from the neighbouring atomic separations increasing. This figure suggests that even if a substantial elastic potential penalty is incurred by this mode it is still likely to be a significant component of the electron-phonon interactions of the system.

The problem of pairing the last two atoms using only z-components in the polarization vectors is rather neatly solved by $\lambda_{8}$. This generator applies vectors of equal magnitude and orientation to atoms 1 and 2, maintaining their separation, however it applies a vector in the opposite direction to atom 3. The opposite signs of the vectors acting on atoms 2 and 3 reduces the $z$-component of their separation. This is not as efficient as the pairing applied to atoms 1 and 2 due to the angle the bond between atoms 2 and 3 makes with the $z$-axis, however the generator also applies a polarization vector to atom 3 which is twice the length as those of the generator $\lambda_{3}$, which compensates.

\begin{figure}
	\includegraphics[width=\columnwidth]{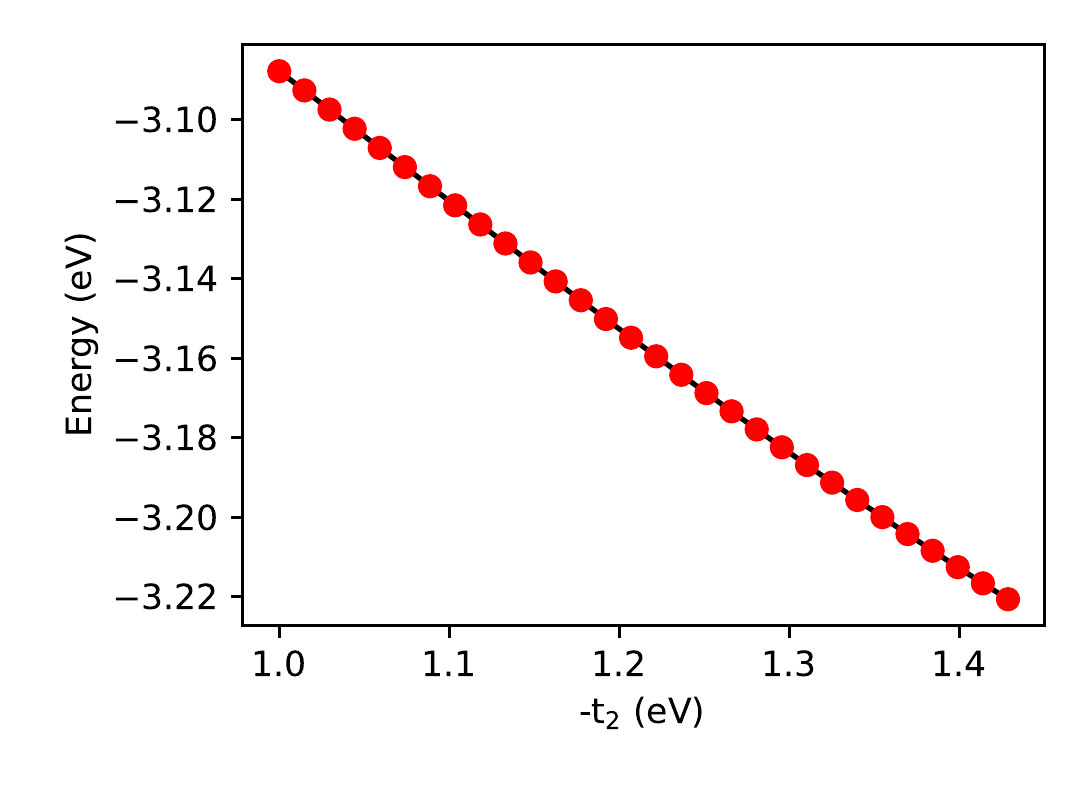}
	\caption{Plot of the energy of a hexagonal Hubbard system in which the second two atoms are paired using polarization vectors with only $z$-components, corresponding to the generator $\lambda_{8}$. The hopping energy between atoms 2 and 3 (denoted by -t$_{2}$) decreases from -1.0 eV to -1.14 eV. The on-site energy $U$ is constant at 5 eV.}
	\label{t2}
\end{figure}

Figure \ref{t2} illustrates the same type of calculation as for the generator $\lambda_{3}$, instead applied to $\lambda_{8}$. Atoms 1, 2 and 3 of the hexagonal system are shifted with eigenvectors corresponding to the $\lambda_{8}$ diagram of Figure \ref{z-modes}. Once again, despite only one separation of the hexagonal system decreasing while slight increases in the separations between atoms 3 and 4, and atoms 6 and 1 occur, the electronic energy of the system decreases, although not to the same extent as for the mode corresponding to $\lambda_{3}$.

As the eigenvectors of these modes are aligned along the z-axis, only the $\gamma^{3}$ terms of Equation \ref{vertex} survive, and as these contain $\sigma^{z}$ matrices acting on the spinor variables, these modes leave the spins unchanged; they are \textit{neutral} modes.

Thus the data of Figures \ref{t1} and \ref{t2} indicates that the Hubbard Hamiltonian, Equation \ref{HubbardH}, is a source of pairing modes, which can be described by the generators $\lambda_{3}$ and $\lambda_{8}$ of an SU(3) Yang-Mills theory.
\begin{figure}
	\includegraphics[width=\columnwidth]{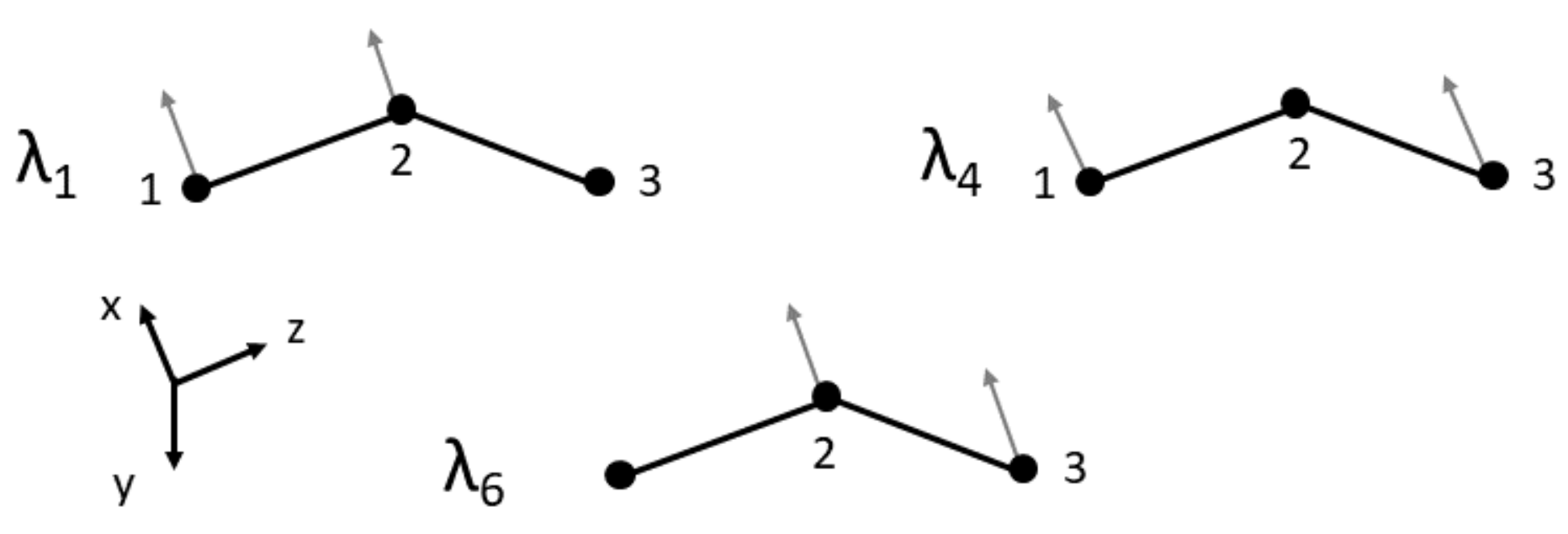}
	\caption{Schematic of the actions of the $\lambda_{1}$, $\lambda_{4}$ and $\lambda_{6}$ generators when applied to the three-atom unit.}
	\label{x-modes}
\end{figure}

However, such pairing modes will only be most effective at lowering the electronic energy of the system if the spins of the hexagonal system can be ordered to give maximum overlap between atoms. This can be achieved using the remaining generators, which due to their orientations, when contracted with the gamma matrices give spin raising and lowering operators.

Figure \ref{x-modes} illustrates the modes $\lambda_{1}, \lambda_{4}$ and $\lambda_{6}$ corresponding to generators which are aligned with the $x$-axis. Each of these contain the same entry, and this multiplies the $\gamma^{1}$ or equivalently $\gamma^{x}$ matrix in the interaction vertex, giving the $\sigma^{x}$ terms of the spin operators $\hat{S}^{\pm} = \frac{1}{\sqrt{2}}(\sigma^{x}\pm i\sigma^{y})$.
\begin{figure}
	\includegraphics[width=\columnwidth]{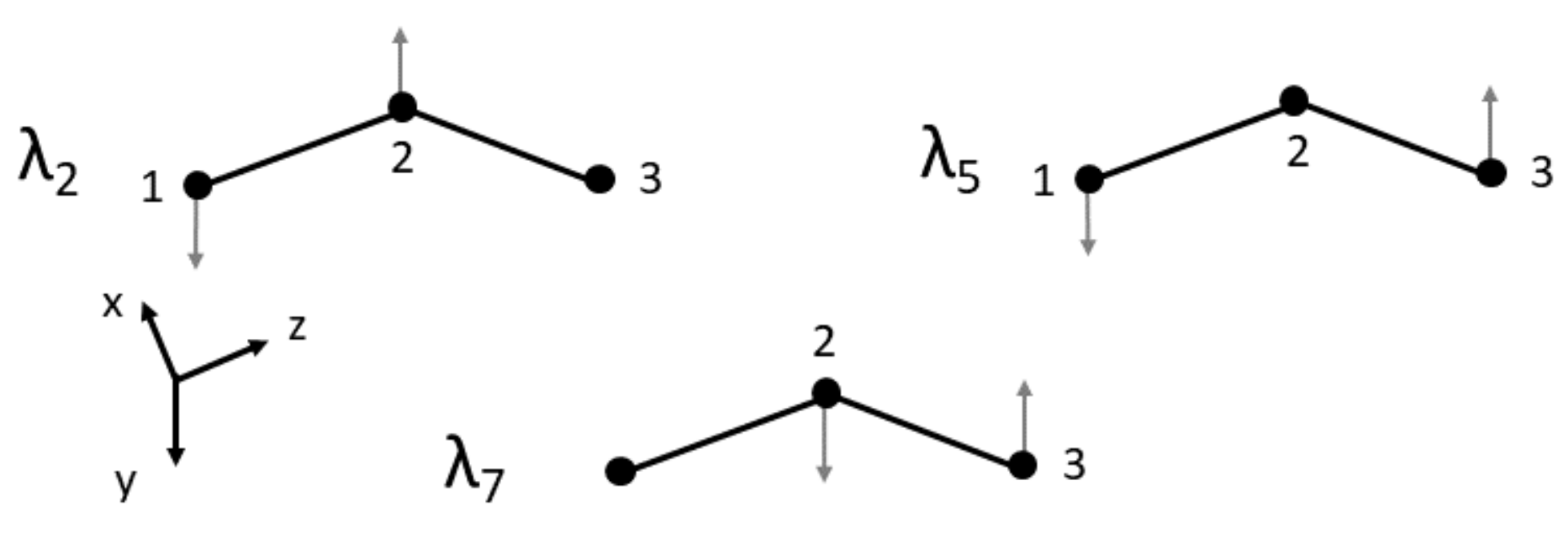}
	\caption{Schematic of the actions of the $\lambda_{2}$, $\lambda_{5}$ and $\lambda_{7}$ generators when applied to the three-atom unit.}
	\label{y-modes}
\end{figure}

Figure \ref{y-modes} illustrates the modes $\lambda_{2}, \lambda_{5}$ and $\lambda_{7}$ which act on the $y$ components of the nuclei positions, which contain the imaginary unit, and a pattern of sign changes which give antiferromagnetic ordering. For example $\lambda_{2}$ applies a polarization vector of $+i$ to atom 1, which when combined with $\lambda_{1}$ gives $\sigma^{x}+i\sigma^{y} = \hat{S}^{+}$ acting on the spinor variables of the electron states. On the hole states, the vertex gives $-\sigma^{x}-i\sigma^{y} = -\hat{S}^{+}$.

On atom 2 the vertex applies a polarization vector of $-i$ which when combined with $\lambda ^{1}$ gives $\sigma^{x}-i\sigma^{y} = \hat{S}^{-}$ acting on the electron states, and $-\sigma^{x}+i\sigma^{y} = -\hat{S}^{-}$ acting on the spinor variables of the hole states. 

This pattern is repeated for the generators $\lambda_{4}$,$\lambda_{5}$, $\lambda_{6}$ and $\lambda_{7}$; combining $\lambda_{4}$ with $\lambda_{5}$ applies $\hat{S}^{+}$ and $-\hat{S}^{+}$ to the electron and hole states of atom 1 respectively, and $\hat{S}^{-}$ and $-\hat{S}^{-}$ to the electron and hole states of atom 3. The combination of $\lambda_{6}$ and $\lambda_{7}$ applies $\hat{S}^{+}$ and $-\hat{S}^{+}$ to atom 2 and $\hat{S}^{-}$ and $-\hat{S}^{-}$ to atom 3.

Of course the oscillatory nature of the fields $\hat{G}^{a}_{\mu}$ will flip the modes between $\hat{S}^{+}$ and $\hat{S}^{-}$ with a time dependence given by the frequency of the mode, however the pattern of alternating spin operations will remain. 

In addition, the combination of $\lambda_{4}$ and $\lambda_{5}$ induces antiferromagnetic ordering on atoms 1 and 3, this combined with the neutral modes $\lambda_{3}$ and $\lambda_{8}$ will facilitate next-to-nearest-neighbor hopping between sites 1 and 3 of the 3 atom sub-unit.

There is a slight subtlety required to give alternating spin raising and lowering operators corresponding the antiferroelectricity. Since these are composite bosons their actions are defined to be of the form:
\begin{equation}
	\hat{G}_{1,2}^{+}|\psi(x_{1})\rangle = (\hat{G}^{1}+i\hat{G}^{2})|\psi(x_{1})\rangle \sim \hat{a}^{\dagger}\epsilon_{x}\hat{a}^{\dagger}\epsilon_{y}|\psi(x_{1})\rangle
\end{equation}
\begin{equation}
	\hat{G}_{1,2}^{-}|\psi(x_{2})\rangle = (\hat{G}^{1}-i\hat{G}^{2})|\psi(x_{2})\rangle \sim -(\hat{a}^{\dagger}\epsilon_{x}\hat{a}^{\dagger}\epsilon_{y})|\psi(x_{2})\rangle
\end{equation}
where $\hat{G}_{1,2}$ is the linear combination of the field operators $G^{1}_{\mu}\lambda^{1}$ and $\hat{G}^{2}\lambda^{2}$, and the imaginary unit and details of the contraction with the gamma matrices are omitted for clarity. Thus although only the $\lambda_{2}$ generator has a minus sign, the operator products for $\hat{G}^{+}/\hat{G}^{-}$ produce polarization vectors in opposite directions, which renders the linear combination $\frac{1}{\sqrt{2}}(\hat{G}^{1}\pm i\hat{G}^{2})$ antiferroelectric. The combinations $\frac{1}{\sqrt{2}}(\hat{G}^{4}\pm i\hat{G}^{5})$ and $\frac{1}{\sqrt{2}}(\hat{G}^{6}\pm i\hat{G}^{7})$ are also antiferroelectric modes for the same reason. 

Thus, expanding the actions of the generators out to see their effects most clearly, starting with $\lambda_{3}$, we have:

\begin{multline}
		\gamma^{\mu}\hat{G}^{3}_{\mu}\lambda_{3}\psi = \\
		\begin{pmatrix}\epsilon_{z}\begin{pmatrix}0&\sigma^{z}\\-\sigma^{z}&0\end{pmatrix}&0&0\\0&-\epsilon_{z}\begin{pmatrix}0&\sigma^{z}\\-\sigma^{z}&0\end{pmatrix}&0\\0&0&0\end{pmatrix}\begin{pmatrix}\psi(x_{1})\\\psi(x_{2})\\\psi(x_{3})\end{pmatrix}
\end{multline}
In this vertex, the $\epsilon_{z}$ polarization vectors act on the positions of the nuclei ($x_{1}$ and $x_{1}$), shifting them in the positive and negative $z$ directions respectively, which gives a Peierls pairing of the two atoms along the $z$-axis. The gamma matrix contains only $\sigma^{z}$ matrices and therefore does not change the $z$ eigenstates of the spin. Identifying the gauge charge with the electron spin, this mode is therefore neutral.

For $\lambda_{8}$ due to the non-zero entries lying on the diagonal, which are contracted with $\gamma^{3}=\gamma^{z}$ we have another neutral mode:
\begin{multline}
	\gamma^{\mu}\hat{G}^{8}_{\mu}\lambda_{8}\psi = \\
	\frac{1}{\sqrt{3}}\begin{pmatrix}\epsilon_{z}\begin{pmatrix}0&\sigma^{z}\\-\sigma^{z}&0\end{pmatrix}&0&0\\0&\epsilon_{z}\begin{pmatrix}0&\sigma^{z}\\-\sigma^{z}&0\end{pmatrix}&0\\0&0&-2\epsilon_{z}\begin{pmatrix}0&\sigma^{z}\\-\sigma^{z}&0\end{pmatrix}\end{pmatrix}\\\times\begin{pmatrix}\psi(x_{1})\\\psi(x_{2})\\\psi(x_{3})\end{pmatrix}
\end{multline}
The $\epsilon_{z}$ polarization vectors shift the nuclei of $\psi(x_{1})$ and $\psi(x_{2})$ in the positive $z$-direction, while shifting the nucleus of $\psi(x_{3})$ twice as far in the negative $z$-direction. This pairs the nuclei at $\psi(x_{2})$ and $\psi(x_{3})$, however it will also facilitate next-nearest-neighbor hopping between sites $x_{1}$ and $x_{3}$.

For the linear combination $\frac{1}{\sqrt{2}}(\lambda_{1}+\lambda_{2})$ we have:
\begin{widetext}
	\begin{multline*}
	\frac{1}{\sqrt{2}}\gamma^{\mu}\big(\hat{G}^{1}_{\mu}\lambda_{1}+\hat{G}^{2}_{\mu}\lambda_{2}\big) = \\
	\begin{pmatrix}0&\frac{1}{\sqrt{2}}\bigg(\epsilon_{x}\begin{pmatrix}0&\sigma^{x}\\-\sigma^{x}&0\end{pmatrix}-\epsilon_{y}\begin{pmatrix}0&-i\sigma^{y}\\i\sigma^{y}&0\end{pmatrix}\bigg)&0\\\frac{1}{\sqrt{2}}\bigg(\epsilon_{x}\begin{pmatrix}0&\sigma^{x}\\-\sigma^{x}&0\end{pmatrix}+\epsilon_{y}\begin{pmatrix}0&i\sigma^{y}\\-i\sigma^{y}&0\end{pmatrix}\bigg)&0&0\\0&0&0\end{pmatrix}\begin{pmatrix}\psi(x_{1})\\\psi(x_{2})\\\psi(x_{3})\end{pmatrix}
	\end{multline*}
\end{widetext}

The $\epsilon_{x}$ and $\epsilon_{y}$ polarization vectors induce antiferroelectricity of the nuclei at $x_{1}$ and $x_{2}$ as per the argument above. However, as the generators are contracted as per: $\gamma^{x}\lambda_{1}$ and $\gamma^{y}\lambda_{2}$. The combinations of the $\sigma^{x}$ and $\sigma^{y}$ matrices from these contractions generate spin raising and lowering operators.

The formation of spin raising and lowering operators for the combinations $\frac{1}{\sqrt{2}}(\lambda_{4}+\lambda_{5})$ and $\frac{1}{\sqrt{2}}(\lambda_{6}+\lambda_{7})$ arises in the same manner, acting on sites 1 and 3, and 2 and 3 respectively.

\section{Discussion}

Thus the preceding argument demonstrates that when viewed as 3$\times$3 linear transformations acting on 3 atom sub-units of 2-dimensional hexagonal materials, the generators of the SU(3) group describe pairing, and spin ordering which lower the electronic energy of the system (as given by the Hubbard Hamiltonian) by facilitating hopping between sites, increasing the admixture of the negative hopping energy in the total ground state energy. 

Therefore, the SU(3) modes above can be considered as proxies for the electron-electron interactions. An interacting liquid of SU(3) modes represents a system of fluctuations which seek to minimize the electron correlations of the system. 

In a previous study\cite{Booth_Wilson_2020} which explored the application of the SU(2) theory to vanadium dioxide, the SU(2) modes, which again were driven by electron-electron interactions, exhibited vacuum expectation values. That is, the quasi-linear system characterized by the SU(2) Yang-Mills theory experiences a phase transition, in which the SU(2) modes describe the changes in position of the metal atoms in the oxide. This allowed the system to be described by an Ising-type Hamiltonian in which the SU(2) modes of adjacent two-atom sites couple with increasing strength as the Temperature is lowered.

However, the hexagonal system is not expected to exhibit such a transformation. The different modes cannot be oriented such that they are orthogonal, for example, $\lambda_{1}$ and $\lambda_{4}$ cannot manifest simultaneously, and perhaps more importantly, neither can the neutral pairing modes $\lambda_{3}$ and $\lambda_{8}$.

Therefore, while the individual modes can combine to lower the electron energy, there is no simple position space configuration which can be adopted that, similarly to the quasi-linear vanadium dioxide, lowers the hopping energy, which as a vacuum expectation value breaks the symmetry. Pairing the atoms along linear chains is a relatively simple symmetry-breaking operation. In hexagonal systems, this is not the case.

There may however be momentum-space configurations which can lower the energy, such as Cooper pairing, however the emergence of such pairing requires a detailed description of the scattering amplitudes of the SU(3) Yang-Mills modes, which as mentioned above will be dealt with in a future study.

This is however a fascinating subject. In recent years remarkable mathematical structures have been found to play unexpected roles in the calculation of scattering amplitudes in planar $\mathcal{N}=4$ Super Yang-Mills theory\cite{Arkani-Hamed2012,Arkani-Hamed2013}. A unification of the formalism of scattering amplitudes in condensed matter and high energy physics in this manner is an attractive theoretical prospect. An immediate simplification in the condensed matter approach developed here is that the generators are intimately related to the directions of the atomic motions in space, and therefore the polarization vectors of the modes become largely redundant.

This presents an exciting opportunity to both examine technologically significant condensed matter systems through the lens of modern scattering amplitude methods, and perhaps provide a more prosaic context for the SU(3) Yang-Mills description of high energy physics which may be more amenable to experiment.

\section{Methods}
The data of Figures \ref{t1} and \ref{t2} were obtained by using the Python package QuSpin\cite{Weinberg2019}. The generators $\lambda_{3}$ and $\lambda_{8}$ were applied to a single hexagonal unit with polarization vectors giving the required hopping energy range. The energies of the system were computed by computing the distances between the atomic sites after the application of the polarization vectors, converting these into hopping energies to give a set of $t_{i}$ where $i$ ranges from $1 \rightarrow 6$, and using exact diagonalization to compute the eigenvalues of the electron Hamiltonian. 
\section{References}
\bibliographystyle{apsrev4-1}
\bibliography{library}

\begin{thebibliography}{13}%
\makeatletter
\providecommand \@ifxundefined [1]{%
 \@ifx{#1\undefined}
}%
\providecommand \@ifnum [1]{%
 \ifnum #1\expandafter \@firstoftwo
 \else \expandafter \@secondoftwo
 \fi
}%
\providecommand \@ifx [1]{%
 \ifx #1\expandafter \@firstoftwo
 \else \expandafter \@secondoftwo
 \fi
}%
\providecommand \natexlab [1]{#1}%
\providecommand \enquote  [1]{``#1''}%
\providecommand \bibnamefont  [1]{#1}%
\providecommand \bibfnamefont [1]{#1}%
\providecommand \citenamefont [1]{#1}%
\providecommand \href@noop [0]{\@secondoftwo}%
\providecommand \href [0]{\begingroup \@sanitize@url \@href}%
\providecommand \@href[1]{\@@startlink{#1}\@@href}%
\providecommand \@@href[1]{\endgroup#1\@@endlink}%
\providecommand \@sanitize@url [0]{\catcode `\\12\catcode `\$12\catcode
  `\&12\catcode `\#12\catcode `\^12\catcode `\_12\catcode `\%12\relax}%
\providecommand \@@startlink[1]{}%
\providecommand \@@endlink[0]{}%
\providecommand \url  [0]{\begingroup\@sanitize@url \@url }%
\providecommand \@url [1]{\endgroup\@href {#1}{\urlprefix }}%
\providecommand \urlprefix  [0]{URL }%
\providecommand \Eprint [0]{\href }%
\providecommand \doibase [0]{http://dx.doi.org/}%
\providecommand \selectlanguage [0]{\@gobble}%
\providecommand \bibinfo  [0]{\@secondoftwo}%
\providecommand \bibfield  [0]{\@secondoftwo}%
\providecommand \translation [1]{[#1]}%
\providecommand \BibitemOpen [0]{}%
\providecommand \bibitemStop [0]{}%
\providecommand \bibitemNoStop [0]{.\EOS\space}%
\providecommand \EOS [0]{\spacefactor3000\relax}%
\providecommand \BibitemShut  [1]{\csname bibitem#1\endcsname}%
\let\auto@bib@innerbib\@empty
\bibitem [{\citenamefont {Novoselov}\ \emph {et~al.}(2012)\citenamefont
  {Novoselov}, \citenamefont {Fal'Ko}, \citenamefont {Colombo}, \citenamefont
  {Gellert}, \citenamefont {Schwab},\ and\ \citenamefont
  {Kim}}]{Novoselov2012}%
  \BibitemOpen
  \bibfield  {author} {\bibinfo {author} {\bibfnamefont {K.~S.}\ \bibnamefont
  {Novoselov}}, \bibinfo {author} {\bibfnamefont {V.~I.}\ \bibnamefont
  {Fal'Ko}}, \bibinfo {author} {\bibfnamefont {L.}~\bibnamefont {Colombo}},
  \bibinfo {author} {\bibfnamefont {P.~R.}\ \bibnamefont {Gellert}}, \bibinfo
  {author} {\bibfnamefont {M.~G.}\ \bibnamefont {Schwab}}, \ and\ \bibinfo
  {author} {\bibfnamefont {K.}~\bibnamefont {Kim}},\ }\href {\doibase
  10.1038/nature11458} {\bibfield  {journal} {\bibinfo  {journal} {Nature}\
  }\textbf {\bibinfo {volume} {490}},\ \bibinfo {pages} {192} (\bibinfo {year}
  {2012})}\BibitemShut {NoStop}%
\bibitem [{\citenamefont {Cao}\ \emph {et~al.}(2018{\natexlab{a}})\citenamefont
  {Cao}, \citenamefont {Fatemi}, \citenamefont {Fang}, \citenamefont
  {Watanabe}, \citenamefont {Taniguchi}, \citenamefont {Kaxiras},\ and\
  \citenamefont {Jarillo-Herrero}}]{Cao2018_2}%
  \BibitemOpen
  \bibfield  {author} {\bibinfo {author} {\bibfnamefont {Y.}~\bibnamefont
  {Cao}}, \bibinfo {author} {\bibfnamefont {V.}~\bibnamefont {Fatemi}},
  \bibinfo {author} {\bibfnamefont {S.}~\bibnamefont {Fang}}, \bibinfo {author}
  {\bibfnamefont {K.}~\bibnamefont {Watanabe}}, \bibinfo {author}
  {\bibfnamefont {T.}~\bibnamefont {Taniguchi}}, \bibinfo {author}
  {\bibfnamefont {E.}~\bibnamefont {Kaxiras}}, \ and\ \bibinfo {author}
  {\bibfnamefont {P.}~\bibnamefont {Jarillo-Herrero}},\ }\href {\doibase
  10.1038/nature26160} {\bibfield  {journal} {\bibinfo  {journal} {Nature}\
  }\textbf {\bibinfo {volume} {556}},\ \bibinfo {pages} {43} (\bibinfo {year}
  {2018}{\natexlab{a}})}\BibitemShut {NoStop}%
\bibitem [{\citenamefont {Cao}\ \emph {et~al.}(2018{\natexlab{b}})\citenamefont
  {Cao}, \citenamefont {Fatemi}, \citenamefont {Demir}, \citenamefont {Fang},
  \citenamefont {Tomarken}, \citenamefont {Luo}, \citenamefont
  {Sanchez-Yamagishi}, \citenamefont {Watanabe}, \citenamefont {Taniguchi},
  \citenamefont {Kaxiras}, \citenamefont {Ashoori},\ and\ \citenamefont
  {Jarillo-Herrero}}]{Cao2018}%
  \BibitemOpen
  \bibfield  {author} {\bibinfo {author} {\bibfnamefont {Y.}~\bibnamefont
  {Cao}}, \bibinfo {author} {\bibfnamefont {V.}~\bibnamefont {Fatemi}},
  \bibinfo {author} {\bibfnamefont {A.}~\bibnamefont {Demir}}, \bibinfo
  {author} {\bibfnamefont {S.}~\bibnamefont {Fang}}, \bibinfo {author}
  {\bibfnamefont {S.~L.}\ \bibnamefont {Tomarken}}, \bibinfo {author}
  {\bibfnamefont {J.~Y.}\ \bibnamefont {Luo}}, \bibinfo {author} {\bibfnamefont
  {J.~D.}\ \bibnamefont {Sanchez-Yamagishi}}, \bibinfo {author} {\bibfnamefont
  {K.}~\bibnamefont {Watanabe}}, \bibinfo {author} {\bibfnamefont
  {T.}~\bibnamefont {Taniguchi}}, \bibinfo {author} {\bibfnamefont
  {E.}~\bibnamefont {Kaxiras}}, \bibinfo {author} {\bibfnamefont {R.~C.}\
  \bibnamefont {Ashoori}}, \ and\ \bibinfo {author} {\bibfnamefont
  {P.}~\bibnamefont {Jarillo-Herrero}},\ }\href {\doibase 10.1038/nature26154}
  {\bibfield  {journal} {\bibinfo  {journal} {Nature}\ }\textbf {\bibinfo
  {volume} {556}},\ \bibinfo {pages} {80} (\bibinfo {year}
  {2018}{\natexlab{b}})}\BibitemShut {NoStop}%
\bibitem [{\citenamefont {Yuan}\ and\ \citenamefont {Fu}(2018)}]{Yuan2018}%
  \BibitemOpen
  \bibfield  {author} {\bibinfo {author} {\bibfnamefont {N.~F.}\ \bibnamefont
  {Yuan}}\ and\ \bibinfo {author} {\bibfnamefont {L.}~\bibnamefont {Fu}},\
  }\href {\doibase 10.1103/PhysRevB.98.045103} {\bibfield  {journal} {\bibinfo
  {journal} {Physical Review B}\ }\textbf {\bibinfo {volume} {98}},\ \bibinfo
  {pages} {1} (\bibinfo {year} {2018})}\BibitemShut {NoStop}%
\bibitem [{\citenamefont {Booth}\ and\ \citenamefont
  {Russo}(2020)}]{Booth2020}%
  \BibitemOpen
  \bibfield  {author} {\bibinfo {author} {\bibfnamefont {J.~M.}\ \bibnamefont
  {Booth}}\ and\ \bibinfo {author} {\bibfnamefont {S.~P.}\ \bibnamefont
  {Russo}},\ }\href {\doibase 10.1038/s41598-020-68958-4} {\bibfield  {journal}
  {\bibinfo  {journal} {Scientific Reports}\ }\textbf {\bibinfo {volume}
  {10}},\ \bibinfo {pages} {12547} (\bibinfo {year} {2020})}\BibitemShut
  {NoStop}%
\bibitem [{\citenamefont {Booth}(2020)}]{Booth_Wilson_2020}%
  \BibitemOpen
  \bibfield  {author} {\bibinfo {author} {\bibfnamefont {J.~M.}\ \bibnamefont
  {Booth}},\ }\href@noop {} {\bibfield  {journal} {\bibinfo  {journal} {arXiv}\
  ,\ \bibinfo {pages} {1901.0719v2}} (\bibinfo {year} {2020})},\ \Eprint
  {http://arxiv.org/abs/1901.07192} {arXiv:1901.07192} \BibitemShut {NoStop}%
\bibitem [{\citenamefont {Dixon}(2013)}]{Dixon2013a}%
  \BibitemOpen
  \bibfield  {author} {\bibinfo {author} {\bibfnamefont {L.~J.}\ \bibnamefont
  {Dixon}},\ }\href {\doibase 10.5170/CERN-2014-008.31} {\  (\bibinfo {year}
  {2013}),\ 10.5170/CERN-2014-008.31},\ \Eprint
  {http://arxiv.org/abs/1310.5353} {arXiv:1310.5353} \BibitemShut {NoStop}%
\bibitem [{\citenamefont {Arkani-Hamed}\ \emph {et~al.}(2017)\citenamefont
  {Arkani-Hamed}, \citenamefont {Huang},\ and\ \citenamefont
  {Huang}}]{Arkani-Hamed2017}%
  \BibitemOpen
  \bibfield  {author} {\bibinfo {author} {\bibfnamefont {N.}~\bibnamefont
  {Arkani-Hamed}}, \bibinfo {author} {\bibfnamefont {T.-C.}\ \bibnamefont
  {Huang}}, \ and\ \bibinfo {author} {\bibfnamefont {Y.-t.}\ \bibnamefont
  {Huang}},\ }\href {http://arxiv.org/abs/1709.04891} {\  (\bibinfo {year}
  {2017})},\ \Eprint {http://arxiv.org/abs/1709.04891} {arXiv:1709.04891}
  \BibitemShut {NoStop}%
\bibitem [{\citenamefont {Arkani-Hamed}\ \emph {et~al.}(2012)\citenamefont
  {Arkani-Hamed}, \citenamefont {Bourjaily}, \citenamefont {Cachazo},
  \citenamefont {Goncharov}, \citenamefont {Postnikov},\ and\ \citenamefont
  {Trnka}}]{Arkani-Hamed2012}%
  \BibitemOpen
  \bibfield  {author} {\bibinfo {author} {\bibfnamefont {N.}~\bibnamefont
  {Arkani-Hamed}}, \bibinfo {author} {\bibfnamefont {J.~L.}\ \bibnamefont
  {Bourjaily}}, \bibinfo {author} {\bibfnamefont {F.}~\bibnamefont {Cachazo}},
  \bibinfo {author} {\bibfnamefont {A.~B.}\ \bibnamefont {Goncharov}}, \bibinfo
  {author} {\bibfnamefont {A.}~\bibnamefont {Postnikov}}, \ and\ \bibinfo
  {author} {\bibfnamefont {J.}~\bibnamefont {Trnka}},\ }\href
  {http://arxiv.org/abs/1212.5605} {\  (\bibinfo {year} {2012})},\ \Eprint
  {http://arxiv.org/abs/1212.5605} {arXiv:1212.5605} \BibitemShut {NoStop}%
\bibitem [{\citenamefont {Arkani-Hamed}\ and\ \citenamefont
  {Trnka}(2013)}]{Arkani-Hamed2013}%
  \BibitemOpen
  \bibfield  {author} {\bibinfo {author} {\bibfnamefont {N.}~\bibnamefont
  {Arkani-Hamed}}\ and\ \bibinfo {author} {\bibfnamefont {J.}~\bibnamefont
  {Trnka}},\ }\href {\doibase 10.1007/JHEP10(2014)030} {\  (\bibinfo {year}
  {2013}),\ 10.1007/JHEP10(2014)030},\ \Eprint {http://arxiv.org/abs/1312.2007}
  {arXiv:1312.2007} \BibitemShut {NoStop}%
\bibitem [{\citenamefont {Zee}(2003)}]{Zee2003}%
  \BibitemOpen
  \bibfield  {author} {\bibinfo {author} {\bibfnamefont {A.}~\bibnamefont
  {Zee}},\ }\href@noop {} {\emph {\bibinfo {title} {{Quantum Field Theory in a
  Nutshell}}}}\ (\bibinfo  {publisher} {Princeton University Press},\ \bibinfo
  {address} {Princeton, New Jersey},\ \bibinfo {year} {2003})\ p.\ \bibinfo
  {pages} {233}\BibitemShut {NoStop}%
\bibitem [{\citenamefont {Pouget}\ \emph {et~al.}(1974)\citenamefont {Pouget},
  \citenamefont {Launois}, \citenamefont {{Rice, Tim}}, \citenamefont
  {Dernier}, \citenamefont {Gossard}, \citenamefont {Villeneuve},\ and\
  \citenamefont {Hagenmuller}}]{Pouget1974}%
  \BibitemOpen
  \bibfield  {author} {\bibinfo {author} {\bibfnamefont {J.~P.}\ \bibnamefont
  {Pouget}}, \bibinfo {author} {\bibfnamefont {H.}~\bibnamefont {Launois}},
  \bibinfo {author} {\bibfnamefont {M.}~\bibnamefont {{Rice, Tim}}}, \bibinfo
  {author} {\bibfnamefont {P.~D.}\ \bibnamefont {Dernier}}, \bibinfo {author}
  {\bibfnamefont {A.}~\bibnamefont {Gossard}}, \bibinfo {author} {\bibfnamefont
  {G.}~\bibnamefont {Villeneuve}}, \ and\ \bibinfo {author} {\bibfnamefont
  {P.}~\bibnamefont {Hagenmuller}},\ }\href@noop {} {\bibfield  {journal}
  {\bibinfo  {journal} {Phys. Rev. B}\ }\textbf {\bibinfo {volume} {10}},\
  \bibinfo {pages} {1801} (\bibinfo {year} {1974})}\BibitemShut {NoStop}%
\bibitem [{\citenamefont {Weinberg}\ and\ \citenamefont
  {Bukov}(2019)}]{Weinberg2019}%
  \BibitemOpen
  \bibfield  {author} {\bibinfo {author} {\bibfnamefont {P.}~\bibnamefont
  {Weinberg}}\ and\ \bibinfo {author} {\bibfnamefont {M.}~\bibnamefont
  {Bukov}},\ }\href {\doibase 10.21468/scipostphys.7.2.020} {\bibfield
  {journal} {\bibinfo  {journal} {SciPost Physics}\ }\textbf {\bibinfo {volume}
  {7}},\ \bibinfo {pages} {1} (\bibinfo {year} {2019})},\ \Eprint
  {http://arxiv.org/abs/1804.06782} {arXiv:1804.06782} \BibitemShut {NoStop}%
\end{thebibliography}%

\end{document}